\documentclass{article}
\usepackage{spconf,amsmath,graphicx}
\usepackage{multirow}
\usepackage{float}
\usepackage{algorithm}
\usepackage{algpseudocode}
\usepackage[fontsize=9pt]{fontsize}
\usepackage{bold-extra}
\usepackage[T1]{fontenc}
\usepackage[subtle]{savetrees}
\usepackage{amssymb}
\usepackage{pifont}
\usepackage{url}
\usepackage{hyperref}

\newcommand{\cmark}{\ding{51}}%
\newcommand{\xmark}{\ding{55}}%

\title{Spell my name: keyword boosted speech recognition}
\name{Namkyu Jung, Geonmin Kim, Joon Son Chung}
\address{Naver Corporation, South Korea}

\begin{document}
%
\maketitle
\begin{abstract}

Recognition of uncommon words such as names and technical terminology is important to understanding conversations in context. However, the ability to recognise such words remains a challenge in modern automatic speech recognition (ASR) systems. 

In this paper, we propose a simple but powerful ASR decoding method that can better recognise these uncommon keywords, which in turn enables better readability of the results. 
The method boosts the probabilities of given keywords in a beam search based on acoustic model predictions. The method does not require any training in advance.

We demonstrate the effectiveness of our method on the LibriSpeeech test sets and also internal data of real-world conversations.
Our method significantly boosts keyword accuracy on the test sets, while maintaining the accuracy of the other words, and as well as providing significant qualitative improvements. 
This method is applicable to other tasks such as machine translation, or wherever unseen and difficult keywords need to be recognised in beam search.

\end{abstract}

\section{Introduction}
\label{sec:intro}

With the advances in deep learning technology, the performance of automatic speech recognition systems have seen tremendous improvements in the recent years~\cite{gulati2020conformer, han2020contextnet, zhang2020pushing, chan2021speechstew}.
The earlier models have struggled to overcome the overfitting problem given insufficient data, but the development of semi-supervised methods like {\em wav2vec 2.0}~\cite{baevski2020wav2vec} has enabled strong performance with a smaller amount of labeled data. Moreover, with the application of various augmentation methods~\cite{park2019specaugment}, models have become better at recognising speech in unfamiliar environments or in noisy environments. 

Despite this, recognising unseen or uncommon words like person names, location names or technical terminologies is left unsolved. In fact, this problem applies to humans as well. No matter how good a person is at listening, it is almost impossible to understand a conversation full of unknown words. Unfortunately, these words might play a very important role in understanding the conversation, even though the total amount of occurrence might be small. Therefore, we focus on the problem of recognising keywords that were not observed during training.


Contextual information has been researched by using fusion methods with trained language models (LM) \cite{gulcehre2015using,shan2019component,stahlberg2018simple}, but also there are works dealing with \textit{contextual biasing} which utilises a specific context such as named entity or personalised contacts. Majority of works require training an additional representation such as bias encoder \cite{pundak18, alon18, jain20}, bias LM~\cite{michaely2016unsupervised}, class based LM~\cite{aleksic2015bringing, le20, kang2020fast}, or additional data augmentation of a named entity with Text-to-speech \cite{zhao2019shallow}. Our work is similar to~\cite{le21} that the bias information is encoded to trie structure. However, they require additional training of RNN-T~\cite{graves12} and LSTM-LM~\cite{merity17} to utilise bias information, whereas our method requires only a list of keywords we are interested in. 

In short, we assume that there is a list of keywords that might appear in the conversation. Our method promotes them through beam search decoding. We use a character-level Connectionist Temporal Classification Model (CTC)~\cite{graves2006connectionist} as an ASR model, specifically a pre-trained {\em wav2vec 2.0} model, but other kinds of model can also be used as long as they are supported by the beam search decoder. We can use the decoding strategy with or without a language model, and in both cases the strategy will prove to be effective. In the real-world,
keywords can be a list of characters in a book, a list of people attending a meeting, or a dictionary of technical terminology related to a lecture.

While performing a beam search, the decoding method favours the given keywords if the input speech is pronounced similar to the given keywords. Since the beam search is based on the acoustic model (AM) probabilities, this method would not be activated unless the speech actually contains a keyword or a word with similar pronunciation. The method does not require any training in advance. 
Our extensive experiments demonstrate that we see a significant boost in keyword accuracy, while maintaining the accuracy on the rest of the output.

\section{Keyword-boosted decoding}
\label{sec:model}

In this section, we describe the proposed keyword-boosted decoding strategy. 

\subsection{Keyword Prefix Tree Search}
We first prepare a prefix tree (Trie) for the given list of keywords. Prefix tree is to figure out if a prefix is a part of some keyword we are paying attention to. Each node in the tree consists of a token and a keyword index if the path from the root node constitutes a keyword, otherwise a non-keyword index (-1). A keyword can be made of multiple words, but to simplify a problem, we only use single-word keywords. 

\begin{figure}[h]
\centering
\includegraphics[scale=0.2]{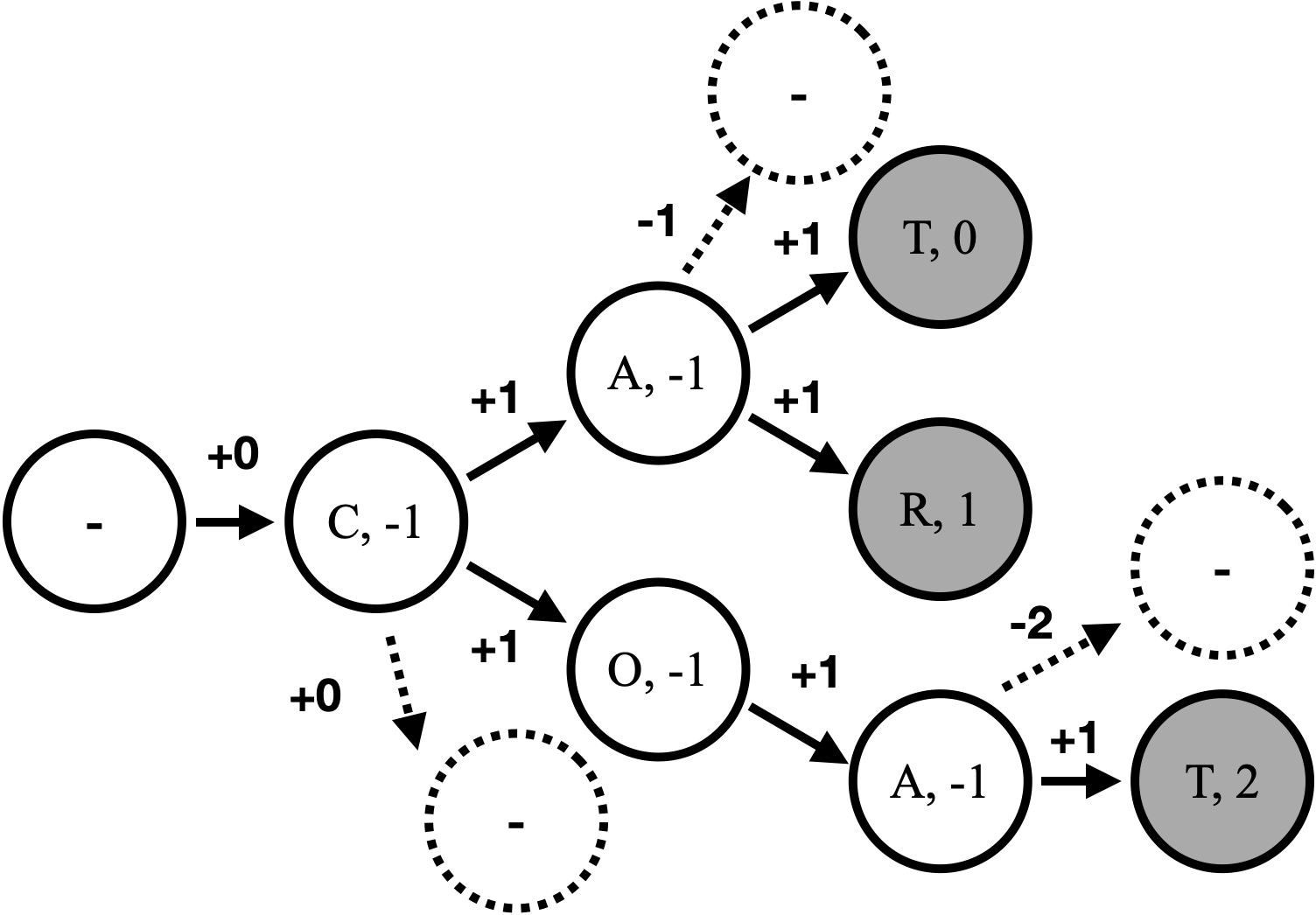}
\caption{Keyword prefix tree when a list of keyword consists of ('cat', 'car', 'coat') in character level.}
\label{fig:trie}
\end{figure}

Figure~\ref{fig:trie} shows an example of keyword prefix tree with sample keyword set \{\textit{cat, car, coat}\}. Each node has its token as a state and a word index. First node represents the root node, $n_0$, and the colored nodes are \textit{leaf\_nodes}. Dashed node means going back to the $n_0$ since there is no child node.

For each step in the beam search, we will find the next node from the corresponding node and the time complexity for finding next node is $O(1)$. If there are many keywords to be considered, building the trees would be relatively time consuming, but this only needs to be done once, and can be reused thereafter in multiple inferences sharing the same keyword list. 

\subsection{Keyword-boosted Beam Search}
Once a prefix tree of keywords $\mathcal{K}$ is prepared, we can decode the acoustic model output with keyword-boosting algorithm. Let $P_{AM}(s)$ for each $s \in V$ be a probability of occurring a token $s$ based on an AM where $V$ is the set of tokens. Beam search proceeds along the time step $t \leq T$ with beam width $B$. For each step $t$, each $b$-th beam has a state $s_{t,b}$, which can be a character or a sub-word. Additionally for each beam, node $n_{t,b}$ of the prefix tree $\mathcal{K}$ which begins at the $n_0$ and \textit{traverse} the tree $\mathcal{K}$ as the beam search proceeds. For CTC models which has \textit{frame-synchronous} decoder~\cite{graves2006connectionist}, $s_{t,b}$ does not always change at every timestep $t$. Therefore, $n_{t,b}$ would be updated only when its state has actually changed.

After each beam search step and the state $s_{t,b}$ is set, the algorithm updates the current tree node $n_{t,b}$ according to the search result (Step 1). If $s_{t,b}$ is in the children of $n_{t,b}$ we can \textit{traverse} $n_{t,b}$ to the child whose state is $s_{t,b}$. If it is not, $n_{t,b}$ should go back to $n_0$. However, there is a chance that $s_{t,b}$ is included in the children of $n_0$. In this case, $n_{t,b}$ will traverse to the child of $n_0$ whose state is $s_{t,b}$.

Before proceeding to the next step $t+1$, the decoder calculates the keyword boosting score $K_{t}$ for each candidate next state like a language model score (Step 2). This score intends to give more weights to the children of the current node which continue to go to the next keyword token. We introduce a new hyperparameter \textit{keyword weight} $w_k$ for controlling the strength of the boosting algorithm. The decoder gives $w_k$ for the children of the $n_{t,b}$ and the 0 for the rest of states. Therefore, the keyword score for each $b$-th beam candidate is 
\[
    K_{t,b}(s)= 
\begin{cases}
    w_k,& \text{if } s \in \text{children}_{\mathbf{K}}(n_{t,b}), n_{t,b} \neq n_0 \\
    0,              & \text{otherwise,}
\end{cases}
\]
for each $s \in V$.

Accordingly, even if an AM does not give much probability to a certain state, $K_{t}$ may give some extra score to boost the state when its node is in the middle of keyword prefix path. Excessively high value of $w_k$ may result in \textit{overboost} (boosting to be a keyword even if it does not actually presented) and a low value may result in nothing but a vanilla beam search. Moreover, this score is not applied from the $n_0$ because we don't want to boost keywords from the beginning, which may also cause overboost. Detailed algorithm for this method is represented in Algorithm~\ref{alg:kb}.

Using the keyword score defined above, for each time $t$, the beam search finds $B$ states in the order of maximising the following score, 
\[ \log{P_{AM}(s)} + w_{LM}\log{P_{LM}}(s|H) + w_k K_t(s|H), \]
where $H$ represents the history.

\begin{algorithm}
\caption{Keyword-boosted beam search step at time $t$.}\label{alg:kb}
\begin{algorithmic}
\State $K_{t,b}(s) \gets K_{t-1,b}(s), \forall s \in V, \forall b \leq B$
\For{$b \leq B$ and $s_{t,b} \neq s_{t-1,b}$}\Comment{update if state has changed.}
    \State \textbf{\texttt{* Step 1: Update current node}}
    \If{$s_{t,b} \in \text{children}_{\mathcal{K}}(n_{t-1, b})$}
        \State $n_{t,b} \gets \text{traverse}_{\mathcal{K}}(n_{t-1, b}, s_{t,b})$
    \ElsIf{$s_{t,b} \in \text{children}_{\mathcal{K}}(n_0)$}
        \State $n_{t,b} \gets \text{traverse}_{\mathcal{K}}(n_0, s_{t,b})$
    \Else
        \State $n_{t,b} \gets n_0$
    \EndIf
    \State \textbf{\texttt{* Step 2: Calculate keyword score}}
    \If{$n_{t,b} \neq n_{t-1,b}$}
        \If{$n_{t,b} = n_0$}
            \State $K_{t,b}(s) \gets 0, \forall s \in V$
        \Else
            \If{$n_{t,b} \neq \textit{leaf\_node}$}
                \State $K_{t,b}(s) \gets -w_k * \text{depth}_{\mathcal{K}}(n_b), \forall s \in V - \{\textit{blank}\}$
                \Comment{subtractive cost}
            \EndIf
            \State $K_{t,b}(s) \gets w_k, \forall s \in  \text{children}_{\mathbf{K}}(n_{t,b})$
        \EndIf
    \EndIf
\EndFor
\end{algorithmic}
\end{algorithm}

\begin{table}
\setlength{\tabcolsep}{5pt}
\centering
\begin{tabular}{c|c|c|c|c} 
\hline
\# keywords                  & dev-clean      & dev-other       & test-clean      & test-other        \\ 
\hline
\hline
1\%          & 714 (1.3)  & 1,008 (2.0)  & 750 (1.4)  & 1,178 (2.3)   \\ 
\hline
5\%         & 2,632 (4.9) & 3,842 (7.5)  & 2,584 (4.9) & 4,126 (7.9)   \\
\hline
\end{tabular}
\caption{The number of keyword occurrence (the ratio to total word occurrence in \%). Keywords account for only a very small percentage of total words.}
\label{table:numkw}
\end{table}

\begin{table*}[ht!]
\centering
\begin{tabular}{c|cc|cc|cc|cc} 
\hline\hline
                                   & \multicolumn{4}{c|}{$w_{LM}=0.0$}                                                 & \multicolumn{4}{c}{$w_{LM}=1.0$}                                                   \\ 
\hline
                                   & \multicolumn{2}{c|}{dev}                & \multicolumn{2}{c|}{test}               & \multicolumn{2}{c|}{dev}                & \multicolumn{2}{c}{test}                 \\ 
\cline{2-9}
                                   & clean              & other              & clean              & other              & clean              & other              & clean              & other               \\ 
\hline
\textbf{\textit{100h fine-tuned} } &                    &                    &                    &                    &                    &                    &                    &                     \\
No Boosting                        & 3.15               & 6.42               & 3.06               & 6.10               & 2.41               & 4.94               & 2.44               & 4.80                \\
1\% Keywords                       & 3.08/3.06          & 6.35/6.33          & 2.96/\textbf{2.94} & 6.00/5.95          & 2.39/\textbf{2.37} & 4.89/4.88          & 2.41/\textbf{2.40} & 4.78/\textbf{4.76}  \\
5\% Keywords                       & 3.06/\textbf{3.01} & 6.28/\textbf{6.27} & 2.95/2.96          & 5.97/\textbf{5.91} & 2.39/2.38          & \textbf{4.86}/4.89 & 2.41/2.42          & 4.78/4.78           \\ 
\hline
\textbf{\textit{960h fine-tuned} } &                    &                    &                    &                    &                    &                    &                    &                     \\
No Boosting                        & 2.16               & 4.56               & 2.13               & 4.46               & 1.77               & 3.51               & 1.78               & 3.61                \\
1\% Keywords                       & 2.13/2.14          & 4.49/4.48          & 2.09/2.10          & 4.39/4.38          & 1.74/1.75          & 3.48/3.48          & 1.76/1.76          & 3.59/\textbf{3.57}  \\
5\% Keywords                       & \textbf{2.09}/2.13 & \textbf{4.42}/4.47 & \textbf{2.08}/2.15 & \textbf{4.37}/4.41 & \textbf{1.72}/1.75 & \textbf{3.43}/3.47 & \textbf{1.75}/1.76 & 3.60/3.61           \\
\hline\hline
\end{tabular}
\caption{Word Error Rates (WER) on LibriSpeech with and without n-gram LM and keyword boosting $w_k=(0.6/1.2)$.}
\label{table:wer}
\end{table*}

\subsection{Keyword subtractive cost}
Being on the tree path does not guarantee that the beam is actually containing a keyword. When the path on $\mathcal{K}$ is about to break because it is not making a keyword, we have to subtract the accumulated value up to the current node. If the next state is not on the tree, we subtract $w_k * \text{depth}_{\mathcal{K}}(n_{t,b})$ as described in Algorithm~\ref{alg:kb}. Note that we do not have to subtract on \textit{CTC blank} state which is still on the keyword path.

In Figure~\ref{fig:trie}, dashed line represents the escape from the tree which receives the subtractive cost and go back to the $n_0$. For example, if the prefix path is \textit{coa} and the next state is going to be \textit{l}, then it will get $-2$ as a subtractive cost and the node will be initialised to the $n_0$.

\section{Experiments}
\label{sec:experiments}

This section describes the ASR model and the datasets used in the experiments, and the results that demonstrate the effectiveness of our proposed method.

\subsection{Baseline ASR system}
\label{subsec:baseline}

We use the {\em wav2vec 2.0} \textsc{Large}~\cite{baevski2020wav2vec} model as the baseline ASR model. The model is pre-trained on the unlabeled audio data of LibriVox dataset~\cite{librivox} and fine-tuned on either 100 hours and 960 hours of transcribed LibriSpeech dataset\cite{panayotov2015librispeech}. The two models are selected to represent scenarios with varying amount of labeled in-domain data. We do not fine-tune the data further, but use a 4-gram word LM that is trained on the LibriSpeech LM corpus~\cite{librispeechlm}. In addition, we use a 6-gram character-level LM during decoding, as a multi-level LM similar to \cite{hori2018end}. This decoding method essentially use the character-level LM, and then substitute character-level probability with word-level ones when each word is made.
We will compare the case of LM weight $w_{LM}=1.0$ or 0.0 (no LM) since the use of LM is optional in our method. 
Beam search decoding has 100 beams and other fixed model weights.

\begin{table}
\centering
\begin{tabular}{c|c|ccc|ccc} 
\hline\hline
       & LM    & \multicolumn{3}{c|}{\xmark} & \multicolumn{3}{c}{\cmark}  \\ 
\hline
       & $w_k$ & P    & R    & F1                           & P    & R    & F1                           \\ 
\hline
\textbf{\textit{100h}}   & 0.0   & 98.9 & 86.0 & 92.0                         & 99.0 & 89.5 & 94.0                         \\
\textbf{\textit{fine}}   & 0.6   & 98.9 & 92.4 & 95.5                         & 98.9 & 92.5 & 95.6                         \\
\textbf{\textit{-tuned}} & 1.2   & 97.5 & 94.7 & 96.1                         & 98.7 & 93.2 & 95.9                         \\ 
\hline
\textbf{\textit{960h}}   & 0.0   & 99.6 & 94.5 & 97.0                         & 99.3 & 95.6 & 97.4                         \\
\textbf{\textit{fine}}   & 0.6   & 99.2 & 97.7 & 98.5                         & 99.2 & 97.9 & 98.5                         \\
\textbf{\textit{-tuned}} & 1.2   & 97.8 & 98.5 & 98.1                         & 98.8 & 98.4 & 98.6                         \\
\hline\hline
\end{tabular}
\caption{Precision (P), Recall (R) and F1-score (F1) on the LibriSpeech test-clean with boosting 1\% keywords for 100h fine-tuned, 960h fine-tuned model respectively.}
\label{table:prf}
\end{table}

\subsection{Datasets and keyword extraction}

\noindent\textbf{LibriSpeech.} 
We evaluate our method on LibriSpeech dev and test sets. On this dataset, we extract the keyword set for each session (book) using the TF-IDF~\cite{papineni01} method. LibriSpeech dataset can be classified by the book where it comes from. From the metadata given by the dataset, we can classify every pair of audio/text file from 960h train dataset into the book name and collect data according to its book. The keywords are only extracted from the training set -- they are not extracted from the dev and test sets since we have to only use keywords obtainable in advance. With the collected text data for each book, we build a TF-IDF model and extract book-wise keywords with top $n$\% of the words according to TF-IDF scores. All single-letter word is excluded in the list. 

We take two set of keywords from book-wise keywords, top 1\% and top 5\% of the words according to TF-IDF scores. The 1\% set has 16.6 words and the 5\% set has 85.5 words for each book on average. We build prefix trees $\mathcal{K}^i$ for each $i$-th book and put them into every audio file from the corresponding book. Since this tree have the whole list of keywords from the book so the most of them are not occurring in each audio file. Table~\ref{table:numkw} presents the total number of extracted keywords in each dataset.\\

\noindent\textbf{In-house dataset.} 
We also perform experiments using Korean-language in-house datasets consisting of \textit{Clova Note} (transcription service) dataset and \textit{NAVER VLive} (Live show for celebrities) dataset. The \textit{Clova Note} dataset has 5,446 audio segments that belong to 365 sessions\footnote{These are purposefully recorded test data, not user-uploaded data.}, and each channel has 7.5 keywords in average extracted by humans. The \textit{NAVER VLive} dataset has 17 audio recordings spoken by two K-pop groups (\textit{BTS}, \textit{Blackpink}), and for each audio, the keyword list contains the members’ name (real name and stage name) of each group, 25 words for \textit{BTS} and 11 words for \textit{Blackpink}. 
The baseline model used in this experiment is also a {\em wav2vec 2.0}-based model, which has been trained on general-domain Korean language data.

\begin{table*}
\centering
\begin{tabular}{l|ll} 
\hline\hline
Positive & keywords     & \textbf{milner},elmwood, sandford, woodley, ojo, dorothy, ozma, scarecrow, miss, lord, tottenhots, pumpkinhead, ...         \\
results  & ground truth & miss \textbf{milner} you shall not leave the house this evening sir                                                         \\
         & $w_k=0$      & miss \textbf{millner} you shall not leave the house this evening sir                                                        \\
         & $w_k=1.2$    & miss \textbf{milner} you shall not leave the house this evening sir                                                         \\ 
\hline
Positive & keywords     & cap'n, boolooroo, button, ghip, trot, \textbf{pinkies}, ghisizzle, blueskins, calder, bill, marianna, angareb, tiggle, ...  \\
results  & ground truth & you are not like my people the \textbf{pinkies} and there is no place for you in our country                                \\
         & $w_k=0$      & you are not like my people the \textbf{pinkeys} and there is no place for you in our country                                \\
         & $w_k=1.2$    & you are not like my people the \textbf{pinkies} and there is no place for you in our country                                \\ 
\hline
Negative & keywords     & servius, praetors, senate, laws, solon, hovel, despotism, julian, decrees, \textbf{athens}, edicts, ...                     \\
results  & ground truth & the worthy friend of \textbf{athanasius} the worthy antagonist of ...                                                       \\
         & $w_k=0$      & the worthy friend of \textbf{athanasius} the worthy antagonist of ..                                                        \\
         & $w_k=1.2$    & the worthy friend of \textbf{athenasius} the worthy antagonist of ...                                                       \\
\hline\hline
\end{tabular}
\caption{Positive and negative samples of transcription on LibriSpeech with keyword boosting or not.}
\label{table:transcription}
\end{table*}

\begin{figure}[h]
\centering
\includegraphics[scale=0.2]{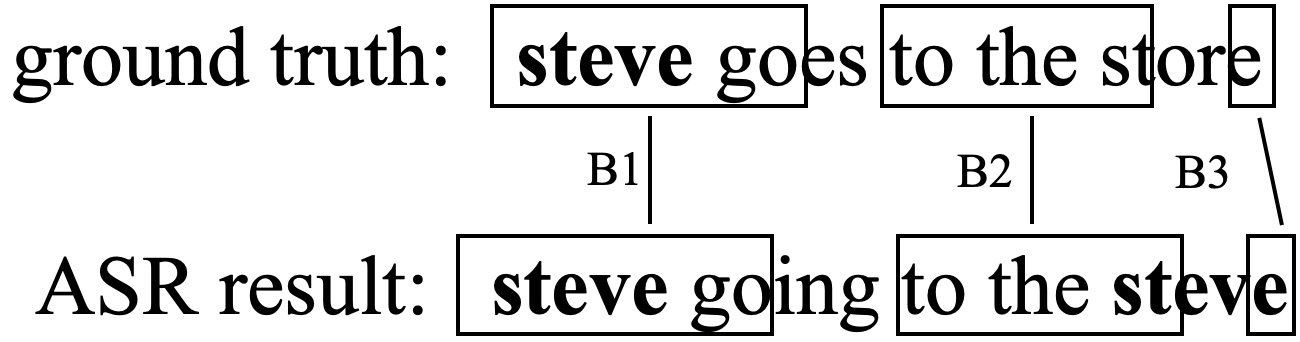}
\caption{Matching blocks between the ground truth and ASR result. Bold text represents a keyword. }
\label{fig:difflib}
\end{figure}

\subsection{Metric}
\label{subsec:metric}
Since the number of keywords is relatively small compared to the size of the corpus (Table~\ref{table:numkw}), the improvement in character error rates (CER) and word error rates (WER) would be limited even if our method is clearly effective in recognising the keywords. 

We use \texttt{python difflib}\footnote{https://docs.python.org/3/library/difflib.html} for string comparison which enables us to compare two strings by matching blocks between them. By comparing between the ground truth and the ASR result with the boosting keywords, we can count the \textit{true positive} (TP): the number of keywords in a matching block, \textit{false positive} (FP): the number of keywords in the ASR result but not in a matching block, \textit{false negative} (FN): the number of keywords in the ground truth but not in a matching block. 

For example in Figure~\ref{fig:difflib} with 3 matching blocks, we have TP equals to $1$ because the keyword \textbf{steve} is in the first matching block (B1) and FP also equals to $1$ because the second \textbf{steve} in ASR result is not inside of any blocks. For a similar reason, FN equals to 0. Consequently, precision is 50\%, recall 100\% and F1-score 66.7\%.

Based on these, we can calculate precision, recall and F1-score with respect to the keywords. Our work focuses on increasing recall while minimizing reduction in precision.




\subsection{Results}
\label{sec:main_results}

\noindent\textbf{LibriSpeech.} 
In Table~\ref{table:wer}, we report the results on LibriSpeech datasets with LM. Upper section is the results with 100h fine-tuned model representing the low-resource environment, and the lower section with 960h fine-tuned model representing full-resource environment. The first row of each section shows the result without keyword boosting, the second row shows the result with boosting top 1\% keywords and the last row with top 5\% keywords. As we discussed in Section~\ref{subsec:metric}, the improvement in WER is relatively small. However in all scenarios, using the boosting method is effective in decreasing WER. Boosting with 5\% of the keywords generally outperforms the other setups.

In addition, we can find that this method is more effective when the model is trained in a low-resource environment in Tables~\ref{table:wer} and \ref{table:prf}. This is an expected observation, because the model with not enough exposure in a target context is probably not familiar with the uncommon words. For a similar reason, the effect is larger without LM than with LM, since the LM should somewhat provide the context and keyword information. Furthermore, high $w_k$ performs better in low-resource environment and moderate $w_k$ is better in the higher-resource environment. 

Table~\ref{table:prf} shows the metrics mentioned in Section~\ref{subsec:metric} on LibriSpeech test-clean. Although the precision is slightly decreased, the gain of recall is significant, resulting in a large gain in the F1-score. The same trend can also be observed in the other test sets. The keyword recall not only depends on $w_k$, but also on $w_{LM}$ although $w_k$ is the more influential factor. Our method is effective on every dataset, but in particular, on test-clean set of LibriSpeech we get a significant improvement of keyword recall from 94.5\% to 98.5\%. The LM often has negative effects on keyword recall in some datasets but it is consistently helpful to 
improving WER in every test scenario, as shown in Table~\ref{table:wer}.\\

\noindent\textbf{In-house dataset.} 
Table~\ref{table:inhouse} shows the character error rate (CER) and the precision-recall values of the experiments done on our in-house datasets. We select larger values of $w_k$ since this is found to be more effective for the Korean dataset. Even though the keywords do not make up a large proportion of the total word occurrences, there is a huge improvement in keyword recall especially from 61.9\% to 82.3\% in the \textit{NAVER VLive} dataset, and a small but consistent improvement in CER. This proves that this method can be effectively utilised for real-world ASR services when the context can be specified.

\subsection{Analysis of side-effects} 
Boosting a larger number of keywords than necessary causes {\em overboost} which makes the results worse for certain types of data. The reduction of precision in Table~\ref{table:prf} show that some results recognise keywords more often than the actual number of occurrence, though this is still relatively rare. We can see an example of this in the last row of Table~\ref{table:transcription}. The keyword \textbf{athens} is very similar to the word in ground truth \textbf{athanasius} so that \textbf{athenasius} has been recognised. Subtractive cost did not work as expected in this example, because the candidates containing the correct word had been pruned in the beam search due to its relatively lower scores than other boosted candidates. This is one of the corner cases that our method was not able to handle.

\begin{table}
\centering
\begin{tabular}{c|c|c|ccc} 
\hline\hline
data  & $w_k$ & CER            & Precision & Recall & F1-score  \\ 
\hline\hline
Clova & 0     & 8.07           & 98.9      & 91.9   & 95.3      \\
Note  & 1     & 7.92           & 98.7      & 93.8   & 96.2      \\
      & 3     & 7.81           & 98.4      & 95.5   & 96.9      \\
      & 5     & \textbf{7.78}  & 98.1      & 96.3   & 97.1      \\
      & 7     & 7.90           & 97.7      & 96.5   & 97.1      \\ 
\hline\hline
NAVER & 0     & 16.74          & 95.8      & 61.9   & 75.2      \\
VLive & 1     & 16.70          & 95.6      & 65.8   & 78.0      \\
      & 3     & 16.62          & 93.8      & 74.0   & 82.7      \\
      & 5     & \textbf{16.58} & 91.7      & 79.9   & 85.4      \\
      & 7     & 16.60          & 87.5      & 82.3   & 84.8      \\
\hline\hline
\end{tabular}
\caption{Result on {\em Clova Note} and {\em VLive} datasets}
\label{table:inhouse}
\end{table}


\section{Conclusions}
\label{sec:conclusions}

We proposed a new keyword-boosted beam search algorithm in speech recognition and demonstrated its performance with keywords extracted from the same book in the LibriSpeech dataset and with human-defined keywords in the in-house dataset. The result shows that this method is clearly helpful for recognising uncommon keywords that are important for understanding the context. We can use this method whenever we lack in-domain training dataset containing difficult keywords, but we only have the list of these words. It does not need any further text data or training process. 

We observed that language models are helpful for improving keyword recall, but our boosting method is far more effective than LM at maintaining lower WER while mitigating the risk of overboost. 

Our method has one requirement: we should have the keyword list in advance. We used TF-IDF on the training data of LibriSpeech to extract book-wise keywords, but there are many ways to obtain keywords in advance, such as using the list of characters in a TV show or the list of terminologies in a lecture.
Strategies to effectively extract keywords are potential areas for future research.

\section{Acknowledgements}

We would like to thank Chan Kyu Lee and Icksang Han for their helpful advice.

\clearpage
\bibliographystyle{IEEEbib}
\bibliography{shortstrings,refs}

\begin{thebibliography}{10}

\bibitem{gulati2020conformer}
Anmol Gulati, James Qin, Chung-Cheng Chiu, Niki Parmar, Yu~Zhang, Jiahui Yu,
  Wei Han, Shibo Wang, Zhengdong Zhang, Yonghui Wu, et~al.,
\newblock ``Conformer: Convolution-augmented transformer for speech
  recognition,''
\newblock {\em arXiv preprint arXiv:2005.08100}, 2020.

\bibitem{han2020contextnet}
Wei Han, Zhengdong Zhang, Yu~Zhang, Jiahui Yu, Chung-Cheng Chiu, James Qin,
  Anmol Gulati, Ruoming Pang, and Yonghui Wu,
\newblock ``Contextnet: Improving convolutional neural networks for automatic
  speech recognition with global context,''
\newblock {\em arXiv preprint arXiv:2005.03191}, 2020.

\bibitem{zhang2020pushing}
Yu~Zhang, James Qin, Daniel~S Park, Wei Han, Chung-Cheng Chiu, Ruoming Pang,
  Quoc~V Le, and Yonghui Wu,
\newblock ``Pushing the limits of semi-supervised learning for automatic speech
  recognition,''
\newblock {\em arXiv preprint arXiv:2010.10504}, 2020.

\bibitem{chan2021speechstew}
William Chan, Daniel Park, Chris Lee, Yu~Zhang, Quoc Le, and Mohammad Norouzi,
\newblock ``Speechstew: Simply mix all available speech recognition data to
  train one large neural network,''
\newblock {\em arXiv preprint arXiv:2104.02133}, 2021.

\bibitem{baevski2020wav2vec}
Alexei Baevski, Henry Zhou, Abdelrahman Mohamed, and Michael Auli,
\newblock ``wav2vec 2.0: A framework for self-supervised learning of speech
  representations,''
\newblock {\em arXiv preprint arXiv:2006.11477}, 2020.

\bibitem{park2019specaugment}
Daniel~S Park, William Chan, Yu~Zhang, Chung-Cheng Chiu, Barret Zoph, Ekin~D
  Cubuk, and Quoc~V Le,
\newblock ``Specaugment: A simple data augmentation method for automatic speech
  recognition,''
\newblock {\em arXiv preprint arXiv:1904.08779}, 2019.

\bibitem{gulcehre2015using}
Caglar Gulcehre, Orhan Firat, Kelvin Xu, Kyunghyun Cho, Loic Barrault, Huei-Chi
  Lin, Fethi Bougares, Holger Schwenk, and Yoshua Bengio,
\newblock ``On using monolingual corpora in neural machine translation,''
\newblock {\em arXiv preprint arXiv:1503.03535}, 2015.

\bibitem{shan2019component}
Changhao Shan, Chao Weng, Guangsen Wang, Dan Su, Min Luo, Dong Yu, and Lei Xie,
\newblock ``Component fusion: Learning replaceable language model component for
  end-to-end speech recognition system,''
\newblock in {\em Proc. ICASSP}. IEEE, 2019, pp. 5361--5635.

\bibitem{stahlberg2018simple}
Felix Stahlberg, James Cross, and Veselin Stoyanov,
\newblock ``Simple fusion: Return of the language model,''
\newblock {\em arXiv preprint arXiv:1809.00125}, 2018.

\bibitem{pundak18}
Pundak Golan, Sainath Tara~N., Prabhavalkar Rohit, Kannan Anjuli, and Zhao
  Ding,
\newblock ``Deep context: End-to-end contextual speech recognition,''
\newblock {\em IEEE Spoken Language Technology Workshop}, 2018.

\bibitem{alon18}
Alon Uri, Pundak Golan, and Sainath Tara~N.,
\newblock ``Contextual speech recognition with difficult negative training
  examples,''
\newblock {\em arXiv preprint arXiv:1810.12170}, 2018.

\bibitem{jain20}
Jain Mahaveer, Keren Gil, Mahadeokar Jay, Zweig Geoffrey, Metze Florian, and
  Saraf Yatharth,
\newblock ``Contextual rnn-t for open domain asr,''
\newblock {\em arXiv preprint arXiv:2006.03411}, 2020.

\bibitem{michaely2016unsupervised}
Assaf~Hurwitz Michaely, Mohammadreza Ghodsi, Zelin Wu, Justin Scheiner, and
  Petar Aleksic,
\newblock ``Unsupervised context learning for speech recognition,''
\newblock in {\em IEEE Spoken Language Technology Workshop}. IEEE, 2016, pp.
  447--453.

\bibitem{aleksic2015bringing}
Petar Aleksic, Mohammadreza Ghodsi, Assaf Michaely, Cyril Allauzen, Keith Hall,
  Brian Roark, David Rybach, and Pedro Moreno,
\newblock ``Bringing contextual information to google speech recognition,''
\newblock 2015.

\bibitem{le20}
Le~Duc, Keren Gil, Chan Julian, Mahadeokar Jay, Fuegen Christian, and Seltzer
  Michael~L.,
\newblock ``Deep shallow fusion for rnn-t personalization,''
\newblock {\em arXiv preprint arXiv:2006.03411}, 2020.

\bibitem{kang2020fast}
Young~Mo Kang and Yingbo Zhou,
\newblock ``Fast and robust unsupervised contextual biasing for speech
  recognition,''
\newblock {\em arXiv preprint arXiv:2005.01677}, 2020.

\bibitem{zhao2019shallow}
Ding Zhao, Tara~N Sainath, David Rybach, Pat Rondon, Deepti Bhatia, Bo~Li, and
  Ruoming Pang,
\newblock ``Shallow-fusion end-to-end contextual biasing.,''
\newblock in {\em Proc. Interspeech}, 2019, pp. 1418--1422.

\bibitem{le21}
Le~Duc, Jain Mahaveer, Keren Gil, Kim Suyoun, Shi Yangyang, Mahadeokar Jay,
  Chan Julian, Shangguan Yuan, Fuegen Christian, Kalinli Ozlem, Saraf Yatharth,
  and Seltzer Michael~L.,
\newblock ``Contextualized streaming end-to-end speech recognition with
  trie-based deep biasing and shallow fusion,''
\newblock {\em Proc. Interspeech}, 2021.

\bibitem{graves12}
Graves Alex,
\newblock ``Sequence transduction with recurrent neural networks,''
\newblock {\em ICML workshop on representation learning}, 2012.

\bibitem{merity17}
Stephen Merity, Nitish~Shirish Keskar, and Richard Socher,
\newblock ``{Regularizing and Optimizing LSTM Language Models},''
\newblock {\em arXiv preprint arXiv:1708.02182}, 2017.

\bibitem{graves2006connectionist}
Alex Graves, Santiago Fern{\'a}ndez, Faustino Gomez, and J{\"u}rgen
  Schmidhuber,
\newblock ``Connectionist temporal classification: labelling unsegmented
  sequence data with recurrent neural networks,''
\newblock in {\em Proc. ICML}, 2006, pp. 369--376.

\bibitem{librivox}
``\url{https://librivox.org/},''
\newblock .

\bibitem{panayotov2015librispeech}
Vassil Panayotov, Guoguo Chen, Daniel Povey, and Sanjeev Khudanpur,
\newblock ``Librispeech: an asr corpus based on public domain audio books,''
\newblock in {\em Proc. ICASSP}. IEEE, 2015, pp. 5206--5210.

\bibitem{librispeechlm}
``\url{https://www.openslr.org/11/},''
\newblock .

\bibitem{hori2018end}
Takaaki Hori, Jaejin Cho, and Shinji Watanabe,
\newblock ``End-to-end speech recognition with word-based rnn language
  models,''
\newblock in {\em IEEE Spoken Language Technology Workshop}. IEEE, 2018, pp.
  389--396.

\bibitem{papineni01}
Papineni Kishore,
\newblock ``{Why inverse document frequency},''
\newblock {\em Proceedings of the North American Association for Computational
  Linguistic}, 2001.

\end{thebibliography}

\end{document}